\documentclass[reprint,nofootinbib]{revtex4-2}
\usepackage[T1]{fontenc}
\usepackage[latin9]{inputenc}
\usepackage[a4paper]{geometry}
\usepackage{babel}
\usepackage{dcolumn}
\usepackage{bm} 
\usepackage{amsmath,amssymb,amsfonts,dsfont}
\usepackage{verbatim}
\usepackage{graphicx,wrapfig,lipsum}
\usepackage{hyperref}
\usepackage{caption}
\usepackage{bbold}
\usepackage{float} 
\usepackage[usenames,dvipsnames]{xcolor}
\usepackage{epsfig}
\usepackage{epstopdf}
\usepackage{tikz}
\usetikzlibrary{calc,shapes.geometric, arrows}
\usepackage{upgreek}
\usepackage{setspace}
\usepackage{enumitem}
\usepackage{array,multirow,bigdelim}
\usepackage{comment}
\usepackage{wasysym}
\usepackage{MnSymbol}
\allowdisplaybreaks

\begin{document}
\widetext

\title{A field-inspired derivation of open string amplitudes}
\author{Humberto Gomez$^{a,b,c}$}
\author{Renann Lipinski Jusinskas$^{b}$}
\author{Sitender Pratap Kashyap$^{b}$}

\affiliation{$^{a}$Facultad de Ciencias Basicas,  Universidad Santiago de Cali,\\
Calle 5 $N^\circ$  62-00 Barrio Pampalinda, Cali, Valle, Colombia}
\affiliation{$^{b}$Institute of Physics of the Czech Academy of Sciences \& CEICO \\ 
Na Slovance 2, 18221 Prague, Czech Republic}
\affiliation{$^{c}$Physics Department, S\~ao Carlos Federal University,\\
Rodovia Washington Lu\'is, km 235, S\~ao Carlos - SP, Brazil}

\begin{abstract}
We propose a new method for computing $N$-point tree-level open string amplitudes, inspired by the field-theoretical perturbiner framework and bypassing world-sheet moduli integration. The construction rests on a weak associativity condition for a point-split, on-shell product of string vertices, introducing a binary-tree prescription with all physical channels manifest. The output is a multi-parametric series representation of the partial amplitude that is amenable to mass-level truncation and, therefore, numerical implementation. Besides the Veneziano amplitude, automatically represented by a dual-channel series with one free parameter, we present its five-point generalization, involving two free parameters and matched against known results in the literature. Higher multiplicities are obtained via the same systematic branching, with no separate input.

\end{abstract}

\maketitle

\section{Introduction}

One of the distinguishing features of string theory concerns scattering amplitudes. Physically speaking, string interactions are not localized in spacetime, a feature that immediately softens the notorious ultraviolet divergences of ordinary quantum field theory, related to short-distance interactions. This smearing appears in the form of world-sheet integrals, and it comes at a price.

In the more than fifty years since the Veneziano amplitude \cite{Veneziano:1968yb}, exact closed-form evaluations of even tree-level integrals in terms of known special functions have remained confined to low multiplicity: a ratio of Gamma functions at four points; generalized hypergeometric functions at five \cite{Kitazawa:1987xj} (see also \cite{Medina:2002nk}) and six points \cite{Oprisa:2005wu}, with limited convergence in the kinematic space. Progress has instead taken the form of an all-orders, algorithmically computable $\alpha'$-expansion  \cite{Broedel:2013aza,Mafra:2016mcc} (see also \cite{Mafra:2022wml} for a detailed review and literature survey). This is a powerful result, but still not resummed into a closed function of the Mandelstam invariants. Only very recently has this picture begun to change, with new closed-form, globally convergent representations \cite{Arkani-Hamed:2024nzc}. This follows a series of both fresh and revisited results related to string scattering at tree \cite{Witten:2013pra,Arkani-Hamed:2019plo,Arkani-Hamed:2022gsa,Eberhardt:2024twy,Saha:2024qpt,Cheung:2025tbr,Arkani-Hamed:2023swr,Arkani-Hamed:2023jwn, Arkani-Hamed:2019mrd} and loop level \cite{Mafra:2017ioj,Mafra:2018qqe,DHoker:2019blr,Claasen:2024ssh,Stieberger:2022lss,Stieberger:2023nol,Eberhardt:2023xck,Eberhardt:2022zay,Arkani-Hamed:2024fyd}.

In this work, we develop a perturbiner-like recursion \cite{Rosly:1996cp,Rosly:1996vr} for open string amplitudes, motivated by the cubic string field theory (SFT) action proposed by Witten \cite{Witten:1985cc}, and its subsequent use in rederiving the Veneziano amplitude \cite{Giddings:1986iy}. Giddings' construction, however, still required integrating over the world-sheet moduli and was not amenable to mass-level truncation. 
We propose a string analogue of the perturbiner recursion combining unintegrated vertex operators via a point-split product that manifestly preserves their relative ordering along the boundary of the world-sheet. The consistency of the nested products completely fixes the relations among the point-splitting parameters entering the recursion, with no integration over world-sheet moduli required at any step. We refer to this feature as a weak associativity condition (WAC), since it is only required up to an overall translation at the boundary.

Our prescription is valid for arbitrary tree-level scattering, both in the type of external string resonances and their number. We focus the presentation on tachyon external states for simplicity, and find that the WAC does far more work than its derivation would suggest. At four points, we derive an alternative representation of the Veneziano amplitude as a dual-channel series depending on a single free parameter, similar in spirit to \cite{Saha:2024qpt}. At five points, with  no input whatsoever from the residues or pole structure of the amplitude, the WAC implies the correct factorization on all physical channels. In addition, it leads to a two-parameter, field theory-like representation of the partial amplitude, which we numerically match with its analogue in \cite{Arkani-Hamed:2024nzc}. This is a non-trivial and, to our knowledge, previously unnoticed relation between a purely algebraic construction and the unitarity of the string amplitude.

\section{$\textrm{Tr}(\phi^3)$ theory as a toy model}\label{sec:trphi3}

We will first motivate our proposal using the classical multi-particle solutions of the $\textrm{Tr}(\phi^3)$ theory, with equation of motion (e.o.m.)\begin{equation}
\Box\phi=\phi^{2}.\label{eq:trphi3-eom}
\end{equation}
The $D$-dimensional d'Alembertian operator is defined as $\Box=\eta^{mn}\partial_{m}\partial_{n}$, where $\eta^{mn}$ is the Minkowski metric (mostly plus signature), which we use to build the scalar product between two vectors. The field $\phi$ is matrix valued and Hermitian. The free solutions
of \eqref{eq:trphi3-eom} are given by single-particle states $\phi_{p}e^{\mathrm{i}k_{p}\cdot x}$, where $\phi_{p}$ is the matrix polarization and $k_{p}^{m}$ is the light-like momentum. The letter $p$ is a single-particle label.

We can build formal multi-particle solutions of the non-linear equation of motion through the ansatz
\begin{equation}
\phi(x)=\sum_{P}\Phi_{P}e^{\mathrm{i}k_{P}\cdot x},\label{eq:scalar-multiparticle}
\end{equation}
where the sum goes over words $P=p_{1}\ldots p_{n}$ with $n\geq1$, and $k_{P}^{m}=k_{p_{1}}^{m}+\ldots k_{p_{n}}^{m}$. This ansatz is a solution of the e.o.m. when the multi-particle matrices $\Phi_{P}$ satisfy
\begin{equation}\label{eq:phi3-recursion-box}
\Box(\Phi_{P}e^{\mathrm{i}k_{P}\cdot x})=\sum_{P=QR}(\Phi_{Q}e^{\mathrm{i}k_{Q}\cdot x})(\Phi_{R}e^{\mathrm{i}k_{R}\cdot x}),
\end{equation}
leading to the algebraic recursion \cite{Mafra:2016ltu,Mafra:2020qst}
\begin{equation}\label{eq:phi3-recursion}
\Phi_{P}=-\frac{1}{k_{P}^{2}}\sum_{P=QR}\Phi_{Q}\Phi_{R},
\end{equation}
The sum is over all possible non-empty sub-word pairs $(Q,R)$ that concatenate to form the word $P$.  This construction is known as the perturbiner (see \cite{LipinskiJusinskas:2026ctz} for a recent review). Note there is an intrinsic order here, as we work with matrix-valued objects ($\Phi_{Q}\Phi_{R} \neq \Phi_{R}\Phi_{Q}$). 

Tree-level partial amplitudes can be computed by attaching an external single-particle state to the multi-particle matrix. This can be expressed as
\begin{multline}
A(1,\ldots,N+1)\\
=-\int \frac{d^{D}x}{(2 \pi)^D}\textrm{Tr}[(\phi_{N+1}e^{\mathrm{i}k_{N+1}\cdot x})\Box(\Phi_{1\ldots N}e^{\mathrm{i}k_{1\ldots N}\cdot x})],\\
=\delta^{D}(k_{1\ldots N}+k_{N+1})\sum_{1\ldots N=QR}\textrm{Tr}(\phi_{N+1}\Phi_{Q}\Phi_{R}),\label{eq:trphi3-partialamplitude}
\end{multline}
matching known results in the literature \cite{Mafra:2020qst}.

Remarkably, each one of the equations above has a direct parallel in the open bosonic string.

\section{Open bosonic string\label{sec:open-string}}

Let us review the basic ingredients from the open bosonic string that we need to establish our proposal. It can be formulated as a two-dimensional conformal field theory (CFT) on the upper half of the complex plane, with conjugate coordinates $z$ and $\bar{z}$, where the real line is the boundary. After the usual doubling trick, we extend the theory to the whole complex plane, but consider only the
holomorphic ($z$-dependent) part of the world-sheet fields. See e.g. \cite{Polchinski:1998rq} for an introduction to this topic.

The gauge fixing of the world-sheet action introduces the Virasoro ghost pair $(b,c)$, yielding
\begin{equation}
    S=\frac{1}{2\pi}\int d^{2}z\left(\frac{1}{\alpha'}\partial X^{m}\bar{\partial}X_{m}+b\bar{\partial}c\right),
\end{equation}
where $\alpha'$ is the string length squared, and $X^m$ denotes the  target-space coordinates. The gauge-fixed action is invariant under the transformations generated by the  Becchi--Rouet--Stora--Tyutin (BRST) charge
\begin{equation}
\mathbf{Q}=\oint(cT-bc\partial c),
\end{equation}
where the energy-momentum tensor $T$, given by
\begin{equation}
T=-\frac{1}{\alpha'}\eta_{mn}\partial X^{m}\partial X^{n}-b\partial c-\partial(bc),
\end{equation}is BRST exact. In terms of the Laurent modes,
\begin{equation}
    \begin{array}{ccc}
\mathbf{L}_{n}=\oint z^{n+1}T(z), &  & \mathbf{b}_{n}=\oint z^{n+1}b(z),\end{array}
\end{equation}
this is expressed as $\{\mathbf{Q},\mathbf{b}_{n}\}=\mathbf{L}_{n}$.

The physical spectrum of the open bosonic string is described by the ghost number one cohomology of $\mathbf{Q}$. The corresponding vertex operators can be written as momentum eigenstates in the form
\begin{equation}
U=cV_{h}e^{\mathrm{i}k\cdot X},\label{eq:vertex-operator}
\end{equation}
where $k^{2}=4(1-h)/\alpha'$. The operator $V_{h}$ has conformal weight $h\in\mathbb{N}_{0}$ and it is built from products of derivatives of $X^{m}$ contracted with tensor polarizations, which are constrained by the BRST closedness condition. For example, the tachyon vertex operator has $V_{0}=1$, while the massless vector has $V_{1}=\varepsilon_{m}\partial X^{m}$, with $k \cdot \varepsilon=0$. Because of the doubling trick, the actual masses of the open string states can be recovered by a rescaling of $\alpha'$ by a factor of $4$.

Finally, the tree-level scattering of open strings is computed via CFT correlators of the vertex operators inserted at the boundary. This will be implicit in our formulas, and automatically introduces a notion of ordering. The partial amplitudes can be cast as
\begin{multline}
A(1,\ldots,N+1) =\bigg\langle U_{N+1}(\infty) \\ \times U_1(1)\int\left(\prod_{i=2}^{N-1}dz_{i}V^{(i)}(z_{i})\right)U_N(0)\bigg\rangle ,\label{eq:open-string-Npt}
\end{multline}
where the ordered integral is expressed as
\begin{equation}
\int\prod_{i=2}^{N-1}dz_{i}=\int_{0}^{1}dz_{N-1}\int_{z_{N-1}}^{1}dz_{N-2}\ldots\int_{z_{3}}^{1}dz_{2},
\end{equation}
and $V=\mathbf{b_{-1}}\cdot U=V_{h}e^{\mathrm{i}k\cdot X}$ is the integrated vertex operator. The partial amplitude is independent of the position of the fixed vertices ($z_{1}=1$, $z_{N}=0$, and $z_{N+1}\to\infty$), which is related to an underlying $\textrm{SL}(2,\mathbb{R})$ residual gauge invariance.

\section{Drawing from SFT}

In order to embed  the partial amplitude \eqref{eq:open-string-Npt} in a recursive framework, like in equation \eqref{eq:trphi3-partialamplitude}, we need a (string) field theory equation of motion.

There is a natural candidate for this equation. The cubic action  for the open bosonic string \citep{Witten:1985cc} yields
\begin{equation}
\mathbf{Q}\Psi=\Psi\star\Psi,\label{eq:cubic-SFT-eom}
\end{equation}
which is a clear parallel to equation \eqref{eq:trphi3-eom}. $\Psi$ is the graded (ghost number one) string field, and the star product ($\star$) is a non-trivial associative operation that captures the string interaction. Geometrically, it outputs an open string field by gluing the halves of two incoming ones, but the precise details will not be relevant here. Note that equation \eqref{eq:cubic-SFT-eom} is invariant under the gauge transformations $\delta\Psi=\mathbf{Q}\Lambda+\Lambda\star\Psi-\Psi\star\Lambda$, with string field parameter $\Lambda$.

Our idea is to propose a simplified  version of the star product, one that works for on-shell building blocks (physical vertex operators) \textit{inside} the CFT correlator that defines the partial amplitude.

First, we introduce the multi-string expansion:
\begin{equation}
\Psi=\sum_{P}\Psi_{P}(z).
\end{equation}
The plane-wave functions, as in the multi-particle expansion \eqref{eq:scalar-multiparticle}, are in-built in the multi-string operators $\Psi_{P}(z)$. The words $P$ have a similar interpretation, with one-letter labels associated to individual, on-shell
vertex operators \eqref{eq:vertex-operator}.

We then define a modified product between two vertex operators as
\begin{equation}
\left(U_{q}\APLstar U_{r}\right)(z)\equiv U_{q}(z+\lambda_{q,r})U_{r}(z-\lambda_{q,r}),\label{eq:new-star-product}
\end{equation}
where the point-splitting parameter, $\lambda_{q,r}$, does not depend on the physical degrees of freedom of the vertices. It makes the right hand side of the equation well defined. Otherwise, as usual in a quantum field theory, the product of two operators at the same point is divergent. Since the multi-string operators are located on the real line, we further demand that the product \eqref{eq:new-star-product} preserves their ordering. In this case, we simply have $\lambda_{q,r}>0$. For the nested product
\begin{multline}
\left(U_{p}\APLstar\left(U_{q}\APLstar U_{r}\right)\right)(z)=U_{p}(z+\lambda_{p,(qr)})\\ \times U_{q}(z-\lambda_{p,(qr)}+\lambda_{q,r})U_{r}(z-\lambda_{p,(qr)}-\lambda_{q,r}),
\end{multline}
the ordering preserving condition ($U_{p}$ to the left of $U_{q}$, to the left of $U_{r}$) is $\lambda_{p,(qr)}>\frac{1}{2}\lambda_{q,r}$. More generally, the point-splitting parameters are labeled by the nested compositions of the $\APLstar$ product, which we denote by $\lambda_{(P),(Q)}$. Additional constraints are related to the associativity of the operation. We demand that
\begin{equation} \label{eq:WAC}
\left(\left(\Psi_{P}\APLstar\Psi_{Q}\right)\APLstar\Psi_{R}\right)(z)=\left(\Psi_{P}\APLstar\left(\Psi_{Q}\APLstar\Psi_{R}\right)\right)(z+\epsilon),
\end{equation}
which implies a \textit{weak} associativity condition, defined up to a translation
$\epsilon$ in $z$, and leading to
\begin{subequations}\label{eq:associativity-constraints}
\begin{align}
\epsilon & =\frac{1}{3}(\lambda_{(P),(QR)}+\lambda_{(PQ),(R)}),\\
\lambda_{(Q),(R)} & =\frac{2}{3}(2\lambda_{(PQ),(R)}-\lambda_{(P),(QR)}),\\
\lambda_{(P),(Q)} & =\frac{2}{3}(2\lambda_{(P),(QR)}-\lambda_{(PQ),(R)}).
\end{align}
\end{subequations}Note that it is not possible to have a vanishing $\epsilon$ without violating the positivity assumption on $\lambda$.

The last ingredient necessary for the recursive computation of the multi-string operators is the inversion of the BRST charge. This is a known construction in SFT, and we need to fix a gauge (see e.g. \cite{Erler:2019vhl} for a review and references therein). We will work with Siegel's gauge, $\mathbf{b}_{0}\cdot\Psi=0$, already assumed in the (single-string) vertex \eqref{eq:vertex-operator}.
The inverse of the BRST charge is given by the operator
\begin{equation}
\mathbf{h}=\frac{\mathbf{b}_{0}}{\mathbf{L}_{0}},
\end{equation}
where $\mathbf{L}_{0}$ is the conformal weight operator. Then it is straightforward to show that $\{\mathbf{Q}, \mathbf{h}\}=\mathds{1}$, which is the known contracting homotopy equation.

Finally, the multi-string operators can be recursively computed via the cubic equation
\begin{equation}
\mathbf{Q}\cdot\Psi_{P}(z)=\sum_{P=QR}\left(\Psi_{Q}\APLstar\Psi_{R}\right)(z),\label{eq:new-eom-cubic}
\end{equation}
which is the string analogue of \eqref{eq:phi3-recursion-box}. By construction, $\Psi_{P}$ is a ghost number one operator, expressed as
\begin{equation}
\Psi_{P}(z)=\sum_{P=QR}\frac{\mathbf{b}_{0}}{\mathbf{L}_{0}}\cdot\left(\Psi_{Q}\APLstar\Psi_{R}\right)(z),\label{eq:multi-string-recursion}
\end{equation}
with the contour defining the modes centered in $z$. 

\section{New prescription}

Our prescription for the alternative computation of the partial amplitude \eqref{eq:open-string-Npt} is given by
\begin{equation}
A(1,\ldots,N+1)=\left\langle U_{N+1}(\infty)\left(\mathbf{Q}\cdot\Psi_{1\ldots N}\right)(z)\right\rangle ,\label{eq:new-Npt-partial}
\end{equation}
where $U_{N+1}(y)$ plays the role of the asymptotic outgoing state ($y\to\infty$), and the incoming states from the asymptotic past ($z\to0$) are described by the multi-string operator $\Psi_{1\ldots N}$. This is the string analogue of equation \eqref{eq:trphi3-partialamplitude}. The weak associativity condition is sufficient within the prescription \eqref{eq:new-Npt-partial}, being consistent, for instance, with gauge invariance and nilpotency of the BRST charge -- see appendix \ref{sec:gaugeinv}.

Let us first consider the three-point partial amplitude. Equation \eqref{eq:new-Npt-partial} leads to
\begin{equation}
A(1,2,3)=\left\langle U_3(\infty)U_{1}(\lambda_{1,2})U_{2}(-\lambda_{1,2})\right\rangle ,
\end{equation}
which coincides with the usual prescription, and is independent of $\lambda_{1,2}$. The more interesting case is the four-point partial amplitude, given by
\begin{multline}
A(1,2,3,4)\\=\left\langle U_4(\infty)\Psi_{12}(z+\lambda_{(12),3})U_3(z-\lambda_{(12),3})\right\rangle \\
+\left\langle U_4(\infty)U_1(z+\lambda_{1,(23)})\Psi_{23}(z-\lambda_{1,(23)})\right\rangle ,\label{eq:new-4pt-partial}
\end{multline}
where we have
\begin{equation}
\Psi_{pq}(z)=\frac{\mathbf{b}_{0}}{\mathbf{L}_{0}}\cdot\left(U_p \APLstar U_q\right)(z).\label{eq:2pt-multi-string}
\end{equation}
The non-trivial step in the construction of the two-string operator  is the evaluation of $\mathbf{L}_{0}$. The expression in parentheses in equation \eqref{eq:2pt-multi-string} is in fact a sum over operators of different weights. Interestingly,  the information on individual conformal weights can be completely read from the powers of $\lambda_{p,q}$. The underlying intuition is easy to grasp. Consider
\begin{equation}
    \mathbf{L}_{0}\cdot e^{\mathrm{i}k\cdot X}(z)=\left(\frac{\alpha' k^{2}}{4}\right)e^{\mathrm{i}k\cdot X}(z),
\end{equation}where the contour defining $\mathbf{L}_{0}$ is around the point $z$. For an operator translated in $z$, we have
\begin{equation}
    \mathbf{L}_{0}\cdot e^{\mathrm{i}k\cdot X}(z+\lambda)=\sum_{n=0}^{\infty}\frac{\lambda^{n}}{n!}\left(\frac{\alpha'k^{2}}{4}+n\right)\partial_{z}^{n}e^{\mathrm{i}k\cdot X}(z),
\end{equation}which is not an eigenvalue equation anymore. The powers of $\lambda$ are associated to the shifted conformal weights of the Taylor expansion. This is expected because $\mathbf{L}_{0}$ does not commute with the generator of world-sheet translations, $\mathbf{L}_{-1}$. In equation \eqref{eq:2pt-multi-string}, we simply replace $\mathbf{L}_{0}$ by a place-holder operator $\ell_{0}(\lambda_{p,q})$, and evaluate it in the last step of the computation.

As a concrete example, let us work with tachyons as external states ($k^{2}=4/\alpha'$), with $U_p=c e^{\mathrm{i} k_p \cdot X}$. Therefore,
\begin{multline}
\Psi_{pq}(z)=\frac{1}{\ell_{0}(\lambda_{p,q})}
\{\lambda_{p,q}[c(z+\lambda_{p,q})+c(z-\lambda_{p,q})]\\ \times e^{\mathrm{i}k_{p}\cdot X(z+\lambda_{p,q})}e^{\mathrm{i}k_{q}\cdot X(z-\lambda_{p,q})}\},
\end{multline}
in which we have used $\mathbf{b}_{0}\cdot\partial c(z)=1$.

After replacing $\Psi_{pq}$ in equation \eqref{eq:new-4pt-partial}, we just need to use the standard tree-level CFT correlators
\begin{equation}
\left\langle c(z_{i})c(z_{j})c(z_{k})\right\rangle =z_{ij}z_{jk}z_{ki},
\end{equation}
where $z_{ij}=z_{i}-z_{j}$, and
\begin{equation}
\left\langle \prod_{j}e^{\mathrm{i}k_{j}\cdot X(z_{j})}\right\rangle =\prod_{i<j}z_{ij}^{\frac{\alpha'}{2}k_{i}\cdot k_{j}},
\end{equation}
with an implicit momentum conserving delta function.

After some algebra, we obtain
\begin{multline}
A(1,2,3,4)\\=\frac{1}{\ell_{0}(\lambda_{1})}\left(\lambda_{1}\right)^{s_{12}-1}\left(1+\frac{\lambda_{1}}{2}\right)^{2-s_{12}-s_{23}}\left(1-\frac{\lambda_{1}}{2}\right)^{s_{23}-2}\\
+\frac{1}{\ell_{0}(\lambda_{2})}\left(\lambda_{2}\right)^{s_{23}-1}\left(1+\frac{\lambda_{2}}{2}\right)^{2-s_{12}-s_{23}}\left(1-\frac{\lambda_{2}}{2}\right)^{s_{12}-2},
\end{multline}
where we denote the generalized Mandelstam variables by $s_{P}=\alpha'k_P^2/4$. The resulting expression only depends on the ratios
\begin{equation}
    \begin{array}{cc}
\lambda_{1}=\frac{\lambda_{1,2}}{\lambda_{(12),3}}, & \lambda_{2}=\frac{\lambda_{2,3}}{\lambda_{1,(23)}}.\end{array}
\end{equation}
This behavior is reproduced for higher points as well. Next, we can make a Taylor expansion for the $\lambda$-ratios, and evaluate $\ell_0$ accordingly, leading to
\begin{multline}\label{eq:4pt-partial-final}
    A(1,2,3,4)=\sum_{N=0}^{\infty}\,\frac{\left(\lambda_{1}\right)^{s_{12}+N-1}}{s_{12}+N-1}C_{N}(s_{12},s_{23})\\+\frac{\left(\lambda_{2}\right)^{s_{23}+N-1}}{s_{23}+N-1}C_{N}(s_{23},s_{12}).
\end{multline}
$C_{N}(a,b)$ is defined as the coefficient of the power $x^N$ in the Taylor expansion of
\begin{equation}
    C(a,b,x)=(1+x/2)^{2-a-b}(1-x/2)^{b-2}.
\end{equation}
Finally, we input the last ingredient in the formula. The WAC, via equation \eqref{eq:associativity-constraints}, implies that
\begin{equation}
\lambda_{2}=\left(\frac{4-2\lambda_1}{2+3\lambda_1}\right).\label{eq:ratio2-4pt}
\end{equation}
Equation \eqref{eq:4pt-partial-final} then matches the Veneziano amplitude for any choice of $\lambda=\lambda_1$, establishing a one-parameter family of representations of the Beta function,
\begin{multline}
B(s-1,u-1;\lambda)=\sum_{N=0}^{\infty}\,\frac{1}{s+N-1}\left(\lambda\right)^{s+N-1}C_{N}(s,u)\\
+\frac{1}{u+N-1}\left(\frac{4-2\lambda}{2+3\lambda}\right)^{u+N-1}C_{N}(u,s),\label{eq:Beta}
\end{multline}
where $s=s_{12}$ and $u=s_{23}$. According to the WAC, we have $0<\lambda<2$. The ratio in \eqref{eq:ratio2-4pt} is in the same range. Note that manifest crossing symmetry ($s\leftrightarrow u$) is achieved when $\lambda=2/3$, which is the optimal value for the series convergence. Considerably more interesting is the fact that this expression is analytic in both $s$ and $u$ everywhere in the kinematic space. This representation can be derived from the analytic continuation of integral definition \eqref{eq:open-string-Npt} at four points -- see appendix \ref{sec:appendixBeta}. We have also checked our prescription with the tachyon scattering by one or two gluons, and the results are displayed in appendix \ref{sec:appendix4pt}.

At five points, equation \eqref{eq:new-Npt-partial} can be decomposed according to the physical channels as
\begin{equation}\label{eq:5pt}
A(1,2,3,4,5)=\sum_{i=1}^5 A_{5}^{(i)},
\end{equation}
where we have
\begin{multline}
        A_{5}^{(1)}=\frac{1}{\ell_{0}(\lambda_{6})}\frac{1}{\ell_{0}(\lambda_{1})}\left(\lambda_{1}\right)^{s_{12}-1}\left(\lambda_{6}\right)^{s_{123}-1}\\\times\left(1-\frac{\lambda_{6}}{2}\right)^{s_{34}-2}
        \left(1+\frac{\lambda_{1}}{2}\right)^{s_{123}-s_{12}-s_{23}+1}\left(1-\frac{\lambda_{1}}{2}\right)^{s_{23}-2}\\\times\left(1+\frac{\lambda_{6}}{2}(1+\lambda_{1})\right)^{s_{23}-s_{123}-s_{234}+1}\\\times\left(1+\frac{\lambda_{6}}{2}(1-\lambda_{1})\right)^{s_{234}-s_{23}-s_{34}+1},
\end{multline}
and similar expressions for the other terms. They are presented in appendix \ref{sec:weak5pt}. There are  eight additional ratios of the point-splitting parameters, given by
\begin{equation}
    \begin{array}{cc}
\lambda_{3}=\frac{\lambda_{2,3}}{\lambda_{(23),4}}, & \lambda_{7}=\frac{\lambda_{1,(23)}}{\lambda_{(1(23)),4}},\\
\lambda_{4}=\frac{\lambda_{3,4}}{\lambda_{2,(34)}}, & \lambda_{8}=\frac{\lambda_{(23),4}}{\lambda_{1,((23)4)}},\\
\lambda_{5}=\frac{\lambda_{3,4}}{\lambda_{(12),(34)}}, & \lambda_{9}=\frac{\lambda_{2,(34)}}{\lambda_{1,(2(34))}},\\
\lambda_{6}=\frac{\lambda_{(12),3}}{\lambda_{((12)3),4}}, & \lambda_{10}=\frac{\lambda_{1,2}}{\lambda_{(12),(34)}}.
\end{array}
\end{equation}  

After we evaluate $\ell_0$ in $A(1,2,3,4,5)$, we obtain the series expansion of the five-point partial amplitude, with all physical poles manifest. Once more, the final ingredient comes from the weak associativity conditions. The resulting ratios can be expressed as
\begin{equation}
 \begin{array}{ccc}
\begin{array}{c}
\lambda_{1}=\frac{2\alpha}{(1-\beta)},\\
\lambda_{5}=2\beta,
\end{array} & \begin{array}{c}
\lambda_{3}=\frac{(2-2\alpha-2\beta)}{(1-\alpha+3\beta)},\\
\lambda_{7}=\frac{(2+6\alpha-2\beta)}{(3+\alpha+5\beta)},
\end{array} & \lambda_{9}=\frac{2(1-\alpha)}{(3\alpha+1)},\end{array}
\end{equation}
and
\begin{equation}\label{eq:factor-condition-5pt}
\lambda_{2i}=\left(\frac{4-2\lambda_{2i-1}}{2+3\lambda_{2i-1}}\right),
\end{equation}
with $i=1,2,3,4,5$. The allowed ranges for the parameters $\alpha$ and $\beta$ are
\begin{equation}\label{eq:triangular-region}
    \begin{array}{ccc}
\alpha>0, & \beta>0, & \alpha+\beta<1.\end{array}
\end{equation}
This triangular region follows from the positivity of the $\lambda$ ratios, and also implies they are smaller than $2$. The weak associativity condition was derived purely from the algebra of the point-split product, with no reference to the residues of the amplitude on physical poles -- see appendix \ref{sec:weak5pt} for the full solution. Yet equation \eqref{eq:factor-condition-5pt} is exactly the condition required for the correct factorization of the five-point amplitude on all its channels. This is a highly non-trivial consistency check: nothing in our method anticipates unitarity, and its emergence here is evidence that the recursion \eqref{eq:multi-string-recursion} captures genuine on-shell string dynamics rather than being a formal analogy.

At six points, equation \eqref{eq:new-Npt-partial} can be naturally organized in 14 terms, which is a direct counting of the binary trees \cite{Mafra:2020qst} generated in the five-string operator $\Psi_{12345}$. Each expression comes with a set of three propagators spanning the physical spectrum. Once the weak associativity condition is imposed, only three independent parameters are left. More generally, this framework directly extends to higher multiplicity. There are $(N-3)$ free parameters describing the $N$-point partial amplitude, which is exactly the dimension of the open-string moduli space. Our proposal is also consistent with the known relation between  scattering trees with cubic vertices (binary trees) and the Catalan numbers. We will explore the combinatorial nature of the construction and other relevant features in an upcoming work. 

\section{Closing remarks and prospects}

We have shown that tree-level open string amplitudes can be computed via an algebraic construction rooted in a cubic equation of motion for multi-string operators, without ever performing a world-sheet moduli integral. Put simply, we dress the usual CFT computations in terms of operator product expansions with a field theory recursion. The essential new ingredient is a weak associativity condition for a point-split  product of string vertices, which promotes the perturbiner construction of section \ref{sec:trphi3} to a field theory of strings. As far as we have observed, the WAC non-trivially implies the correct factorization of the amplitude.

The output of our prescription is a series expansion with all the physical channels of the amplitude manifest, amenable to level truncation. This is in contrast with earlier results in the literature, in which the pole expansions of disk integrals did not yield everywhere-convergent series \cite{Hopkinson:1969er,Zakrzewski:1969lzk}. There is also a strong suggestion that the multi-parametric freedom we found is connected to an associahedron interpretation, and we plan to report on this soon. 

At four points, we have tested equation \eqref{eq:Beta} for different values of $s_{12}$ and $s_{23}$, comparing them against the exact Beta function and against the structurally different, but likewise one-parameter, expansion of  \cite{Saha:2024qpt}. Outside the physical region, we recover, for example, the notable result $B(1/2,1/2)=\pi$. With $\lambda=2/3$, the representation \eqref{eq:Beta} has a ten decimal point agreement for a cut-off at $N=16$. In the physical region, convergence is also achieved with mass-level truncation. For instance, random scans of the zeros of the Beta function $s+u=-Z$, with $Z=10,20,30$, are reproduced to $0.00001$ accuracy for mass-level truncations at $N=34,54,77$, respectively.

At five points, we have numerically compared the generating series \eqref{eq:5pt} against the closed-form representation of \cite{Arkani-Hamed:2024nzc}. Once more, we have verified convergence both inside and outside the physical kinematic region. For example, consider the asymmetric physical-region point ($s_{12}=-1.2$, $s_{23}=-2.3$, $s_{34}=-3.4$, $s_{123}=-1.23$, $s_{234}=-2.34$). Choosing $\alpha=\beta=1/3$, agreement with \cite{Arkani-Hamed:2024nzc} reaches nine digit precision already at $N=50$. For $\alpha=1/9$, $\beta=1/6$ we reach six digits at the same cut-off.

Attached to the arXiv submission, we have a Mathematica notebook implementing our formulas, and those of references \cite{Saha:2024qpt} and \cite{Arkani-Hamed:2024nzc}, all adapted to our conventions for a side-by-side comparison. We encourage the reader to have a look and freely play with the parameters in order to test them. Note that the cut-off required for convergence generally grows larger near the boundary of the allowed region \eqref{eq:triangular-region}, and convergence is not guaranteed throughout the full parameter range. A necessary condition for the underlying Taylor expansions to be well defined is
\begin{equation}\label{eq:convergence-necessary}
\alpha < 2\beta, \qquad \beta < 2\alpha,
\end{equation}
corresponding to a triangle with vertices $(0,0)$, $(1/3,2/3)$, $(2/3,1/3)$, strictly contained within the region \eqref{eq:triangular-region}. The two triangles share the vertex $(0,0)$ and part of the hypotenuse $\alpha+\beta=1$, with the smaller triangle covering exactly one third of the original area. Outside this sub-region, the truncated series may fail to converge for some choices of external kinematics, even while \eqref{eq:triangular-region} is satisfied. A complete characterization of the kinematics-dependent convergence domain is left for future work. Away from these singular regions of the kinematic space, our series expansions constitute a reliable tool for precision computation, which we hope will complement the recent framework of \cite{Eberhardt:2024twy}.

Beyond the open bosonic string, our method should directly apply to the open spinning string (Ramond--Neveu--Schwarz formalism) with appropriate picture-changing operator insertions. An extension to the pure spinor superstring is also plausible, especially given how advanced its all-multiplicity, tree-level formulation is \cite{Mafra:2010jq,Mafra:2011nv,Mafra:2011nw,Mafra:2014oia}. Operator ordering is a core ingredient here, and the generalization of the WAC to the closed (super) string is not immediate. While the Kawai--Lewellen--Tye \cite{Kawai:1985xq} relations provide a simple construction of closed string amplitudes in terms of the open string ones, it would be desirable to have explicit formulas more aligned with our series expansions. String-like models could also benefit from our framework, such as a first-principles derivation of the amplitude prescription in the asymmetrically twisted string \cite{Jusinskas:2021bdj}. A more ambitious, considerably less immediate direction is the extension of our framework to one loop, with an appropriate annulus generalization of the weak associativity condition and the sum over an infinite number of virtual states to be integrated in the loop.

\begin{acknowledgments}
We thank Carlos Mafra, Sebastian Mizera, and Oliver Schlotterer for valuable comments and suggestions. HG is partially supported by FAPESP, grant 2026/00820-8. RLJ is supported by the GA\v{C}R grant 25-16244S from the Czech Science Foundation. SPK is supported by the Marie Sk\l odowska--Curie Actions -- COFUND project, which is co-funded by the European Union (Physics for Future -- Grant Agreement No. 101081515).
\end{acknowledgments}

\bibliography{refs}


\appendix

\section{Gauge invariance of the prescription} \label{sec:gaugeinv}

Let us reconsider the partial amplitude \eqref{eq:new-Npt-partial}, which can be rewritten using equation \eqref{eq:multi-string-recursion} as
\begin{equation}
    A=\sum_{1\ldots N=QR}\left\langle U_{N+1}(\infty)\left(\Psi_{Q}\APLstar\Psi_{R}\right)(z)\right\rangle,
\end{equation}
with short-hand $A=A(1,\ldots,N+1)$. The gauge transformation on the external leg $(N+1)$ is given by a BRST-exact operator, $\delta U_{N+1}=\mathbf{Q}\cdot \Lambda$. Therefore,
\begin{equation}
    \delta A 
=\sum_{1\ldots N=QR}\left\langle \mathbf{Q}\cdot\Lambda(\infty)\left(\Psi_{Q}\APLstar\Psi_{R}\right)(z)\right\rangle .
\end{equation}

The contour defining the action of the BRST charge can be deformed in such a way that
\begin{multline}
    \delta A =-\sum_{1\ldots N=QR}\left\langle \Lambda(\infty)\mathbf{Q}\cdot\left(\Psi_{Q}\APLstar\Psi_{R}\right)(z)\right\rangle ,\\=\sum_{1\ldots N=QR} \big[\left\langle \Lambda(\infty)\left(\Psi_{Q}\APLstar \mathbf{Q}\cdot\Psi_{R}\right)(z)\right\rangle \\ -\left\langle \Lambda(\infty)\left(\mathbf{Q}\cdot\Psi_{Q}\APLstar\Psi_{R}\right)(z)\right\rangle\big].
\end{multline}
which is similar to an integration by parts. Now using equation \eqref{eq:new-eom-cubic}, we obtain
\begin{multline}
    \delta A = \sum_{1\ldots N=QR} \big[\left\langle \Lambda(\infty)\left(\Psi_{Q}\APLstar\left(\Psi_{R}\APLstar\Psi_{S}\right)\right)(z)\right\rangle  \\ -\left\langle \Lambda(\infty)\left(\left(\Psi_{Q}\APLstar\Psi_{R}\right)\APLstar\Psi_{S}\right)(z)\right\rangle\big].
\end{multline}
Because of the WAC, the first and the second lines cancel each other. This is possible because $\Lambda$ is inserted at infinity. If we want to consider gauge transformations of the incoming external legs \{1,\ldots,N\}, the analysis is less trivial because it involves the construction of a multi-string gauge parameter. The procedure mimics the field theory construction (see \cite{LipinskiJusinskas:2026ctz}), but we do not add details here.

\section{Beta function from the integral representation} \label{sec:appendixBeta}

Our representation for the Beta function can be derived from an analytic continuation of its defining integral. With a simple split in the integration domain, we obtain
\begin{align}
    B(s,u)
    &=\int_{0}^{\frac{2\lambda}{2+\lambda}}y^{s-1}(1-y)^{u-1}dy\nonumber\\
    &+\int_{\frac{2\lambda}{2+\lambda}}^{1}y^{s-1}(1-y)^{u-1}dy,
\end{align}
introducing the parameter $0<\lambda<2$ that appears on the right hand side of equation \eqref{eq:Beta}. We assume $s>0$ and $u>0$, so the integration is finite. 

Next, we make a coordinate transformation $y=2z/(2+z)$ in the first integral, and  $y=1-2z/(2+z)$ in the second integral, leading to
\begin{multline}
    B(s,u;\lambda)=\int_{0}^{\lambda}z^{s-1}(1+z/2)^{-s-u}(1-z/2)^{u-1}dz\\+\int_{0}^{\frac{4-2\lambda}{2+3\lambda}}z^{u-1}(1+z/2)^{-s-u}(1-z/2)^{s-1}dz.
\end{multline}

In order to solve the integrals, we make a Taylor expansion around $z=0$, precisely yielding equation \eqref{eq:Beta}. The last step is the analytical continuation beyond the convergence region of the integral.

\section{Other four-point computations} \label{sec:appendix4pt}

Here we verify our prescription \eqref{eq:new-Npt-partial} for four-point amplitudes involving tachyons and one or two gluons. The vertex operators of the latter are given by $U_\textrm{g}=c (\varepsilon_m \partial X^m ) e^{\mathrm{i}k\cdot X}$, with $k^2=k \cdot\varepsilon=0$.

The partial amplitude with one gluon will be denoted by $A(1,2,3,4_{\textrm{g}})$. In equation \eqref{eq:new-4pt-partial}, we simply replace $U_4(\infty)$ by the gluon vertex operator.  In practice, we use $U_4(y)$ and only take the limit $y \to \infty$ later. For BRST-closed states, the divergent pieces in $y$ are proportional to on-shell vanishing quantities, and the final expression follows the recurrent pattern in terms of $\lambda$-ratios. We then obtain
\begin{multline}
    A(1,2,3,4_{\textrm{g}})= -(\mathrm{i}\alpha'/2)\\\times\Big\{\frac{1}{\ell_{0}(\lambda_{1})}\left(\lambda_{1}\right)^{s_{12}-1}\left(1+\frac{\lambda_{1}}{2}\right)^{1-s_{12}-s_{23}}\left(1-\frac{\lambda_{1}}{2}\right)^{s_{23}-2}\\\times\Big[(k_{1}\cdot\varepsilon_{4})\left(1+\frac{\lambda_{1}}{2}\right)+(k_{2}\cdot\varepsilon_{4})\left(1-\frac{\lambda_{1}}{2}\right)\Big]\\-\frac{1}{\ell_{0}(\lambda_{2})}\left(\lambda_{2}\right)^{s_{23}-1}\left(1+\frac{\lambda_{2}}{2}\right)^{1-s_{12}-s_{23}}\left(1-\frac{\lambda_{2}}{2}\right)^{s_{12}-2}\\\times\Big[(k_{2}\cdot\varepsilon_{4})\left(1-\frac{\lambda_{2}}{2}\right)+(k_{3}\cdot\varepsilon_{4})\left(1+\frac{\lambda_{2}}{2}\right)\Big]\Big\},
\end{multline}
which can be recast in terms of the Beta function as
\begin{multline}
    A(1,2,3,4_{\textrm{g}})=-\frac{\alpha'}{2}\big[(k_{1}\cdot\varepsilon_{4})B(s_{12},s_{23};\lambda)\\+(k_{2}\cdot\varepsilon_{4})B(s_{12},s_{23}+1;\lambda)\big],
\end{multline}
after using the WAC, momentum conservation, and the transversality of the gluon polarization.

The computation of the partial amplitude involving two tachyons and two gluons is slightly longer, but can finally be expressed as\begin{widetext}
\begin{multline}
A(1,2,3_{\textrm{g}},4_{\textrm{g}})=\left(\frac{\alpha^{\prime}}{2}\right)^{2}\Big\{\frac{1}{\ell_{0}(\lambda_{1})}\left(\lambda_{1}\right)^{s_{12}-1}\left(1+\frac{\lambda_{1}}{2}\right)^{1-s_{12}-s_{23}}\left(1-\frac{\lambda_{1}}{2}\right)^{s_{23}-1}\Big[-\left(\frac{2}{\alpha^{\prime}}\right)(\varepsilon_{3}\cdot\varepsilon_{4})+(k_{1}\cdot\varepsilon_{3})(k_{1}\cdot\varepsilon_{4})\\+(k_{2}\cdot\varepsilon_{3})(k_{2}\cdot\varepsilon_{4})+(k_{1}\cdot\varepsilon_{3})(k_{2}\cdot\varepsilon_{4})\left(1+\frac{\lambda_{1}}{2}\right)^{-1}\left(1-\frac{\lambda_{1}}{2}\right)+(k_{2}\cdot\varepsilon_{3})(k_{1}\cdot\varepsilon_{4})\left(1+\frac{\lambda_{1}}{2}\right)\left(1-\frac{\lambda_{1}}{2}\right)^{-1}\Big]\\+\frac{1}{\ell_{0}(\lambda_{2})}\left(\lambda_{2}\right)^{s_{23}}\left(1+\frac{\lambda_{2}}{2}\right)^{1-s_{12}-s_{23}}\left(1-\frac{\lambda_{2}}{2}\right)^{s_{12}-2}\Big[-\left(\frac{2}{\alpha'}\right)\left(\varepsilon_{3}\cdot\varepsilon_{4}\right)+(k_{1}\cdot\varepsilon_{3})(k_{1}\cdot\varepsilon_{4})\\+(k_{2}\cdot\varepsilon_{3})(k_{2}\cdot\varepsilon_{4})+(k_{1}\cdot\varepsilon_{3})(k_{2}\cdot\varepsilon_{4})\lambda_{2}\left(1+\frac{\lambda_{2}}{2}\right)^{-1}+(k_{2}\cdot\varepsilon_{3})(k_{1}\cdot\varepsilon_{4})\left(\lambda_{2}\right)^{-1}\left(1+\frac{\lambda_{2}}{2}\right)\Big]\Big\}
\end{multline}
\end{widetext}
We are intentionally leaving $A(1,2,3,4_{\textrm{g}})$ and $A(1,2,3_{\textrm{g}},4_{\textrm{g}})$ in an extended form in order to help visualize our method. Once we evaluate $\ell_0$, we have two series with the physical poles manifest and coefficients involving the polarizations of the gluons and the external momenta. Gauge invariance of both amplitudes can be easily checked using the traditional representations in terms of  gamma functions.

\section{Five-point expressions and WAC} \label{sec:weak5pt}

The remaining terms of the five-point partial amplitude \eqref{eq:5pt} are given by
\begin{widetext}
    \begin{multline}
        A_{5}^{(2)}=\frac{1}{\ell_0(\lambda_{7})}\frac{1}{\ell_{0}(\lambda_{2})}\left(\lambda_{2}\right)^{s_{23}-1}\left(\lambda_{7}\right)^{s_{123}-1}\left(1+\frac{\lambda_{2}}{2}\right)^{s_{123}-s_{12}-s_{23}+1}\left(1-\frac{\lambda_{2}}{2}\right)^{s_{12}-2}\left(1+\frac{\lambda_{7}}{2}\right)^{s_{23}+1-s_{123}-s_{234}}\\\times\left(1-\frac{\lambda_{7}}{2}(1-\lambda_{2})\right)^{s_{234}+1-s_{23}-s_{34}}\left(1-\frac{\lambda_{7}}{2}(1+\lambda_{2})\right)^{s_{34}-2},
    \end{multline}
    \begin{multline}
        A_{5}^{(3)}=\frac{1}{\ell_{0}(\lambda_{8})}\frac{1}{\ell_{0}(\lambda_{3})}\left(\lambda_{3}\right)^{s_{23}-1}\left(\lambda_{8}\right)^{s_{234}-1}\left(1+\frac{\lambda_{3}}{2}\right)^{s_{234}-s_{23}-s_{34}+1}\left(1-\frac{\lambda_{3}}{2}\right)^{s_{34}-2}\left(1+\frac{\lambda_{8}}{2}\right)^{s_{23}+1-s_{123}-s_{234}}\\\times\left(1-\frac{\lambda_{8}}{2}(1+\lambda_{3})\right)^{s_{12}-2}\left(1-\frac{\lambda_{8}}{2}(1-\lambda_{3})\right)^{s_{123}-s_{12}-s_{23}+1},
    \end{multline}
    \begin{multline}
        A_{5}^{(4)}=\frac{1}{\ell_{0}(\lambda_{9})}\frac{1}{\ell_{0}(\lambda_{4})}\left(\lambda_{4}\right)^{s_{34}-1}\left(\lambda_{9}\right)^{s_{234}-1}\left(1+\frac{\lambda_{4}}{2}\right)^{s_{234}-s_{23}-s_{34}+1}\left(1-\frac{\lambda_{4}}{2}\right)^{s_{23}-2}\left(1-\frac{\lambda_{9}}{2}\right)^{s_{12}-2}\\\times\left(1+\frac{\lambda_{9}}{2}(1+\lambda_{4})\right)^{s_{23}+1-s_{123}-s_{234}}\left(1+\frac{\lambda_{9}}{2}(1-\lambda_{4})\right)^{s_{123}-s_{12}-s_{23}+1},
    \end{multline}
    \begin{multline}
       A_{5}^{(5)}=\frac{1}{\ell_{0}(\lambda_{10})}\frac{1}{\ell_{0}(\lambda_{5})}\left(\lambda_{10}\right)^{s_{12}-1}\left(\lambda_{5}\right)^{s_{34}-1}\left(1+\frac{1}{2}(\lambda_{10}+\lambda_{5})\right)^{s_{23}+1-s_{123}-s_{234}}\left(1+\frac{1}{2}(\lambda_{10}-\lambda_{5})\right)^{s_{123}-s_{12}-s_{23}+1}\\\times\left(1-\frac{1}{2}(\lambda_{10}+\lambda_{5})\right)^{s_{23}-2}\left(1-\frac{1}{2}(\lambda_{10}-\lambda_{5})\right)^{s_{234}-s_{23}-s_{34}+1},
    \end{multline}
\end{widetext}

In this computation, we need to determine the WAC on the product of four operators. They translate to
\begin{multline}\label{eq:WAC-5pt}
    \left(\left(\left(U_{1}\APLstar U_{2}\right)\APLstar U_{3}\right)\APLstar U_{4}\right)(z)\\=\left(\left(U_{1}\APLstar\left(U_{2}\APLstar U_{3}\right)\right)\APLstar U_{4}\right)(z+\epsilon_{1}),\\=\left(U_{1}\APLstar\left(\left(U_{2}\APLstar U_{3}\right)\APLstar U_{4}\right)\right)(z+\epsilon_{2}),\\=\left(U_{1}\APLstar\left(U_{2}\APLstar\left(U_{3}\APLstar U_{4}\right)\right)\right)(z+\epsilon_{3}),\\=\left(\left(U_{1}\APLstar U_{2}\right)\APLstar\left(U_{3}\APLstar U_{4}\right)\right)(z+\epsilon_{4}).
\end{multline}

The point-splitting parameters are $\lambda_{1,2}$, $\lambda_{2,3}$, $\lambda_{3,4}$, $\lambda_{(12),3}$, $\lambda_{1,(23)}$, $\lambda_{(23),4}$, $\lambda_{2,(34)}$, $\lambda_{((12)3),4}$, $\lambda_{(1(23)),4}$, $\lambda_{1,((23)4)}$, $\lambda_{1,(2(34))}$, and $\lambda_{(12),(34)}$. However, only three of them are independent after imposing the WAC implied by \eqref{eq:WAC-5pt}. We choose $\lambda_{(12),(34)}$ and
\begin{equation}
\begin{array}{cc}
       \frac{\lambda_{1,2}}{\lambda_{(12),(34)}}\equiv2\alpha,&\frac{\lambda_{3,4}}{\lambda_{(12),(34)}}\equiv2\beta.
\end{array}
\end{equation}
The remaining ones are then expressed as
\begin{align}
\lambda_{2,3}&=(1-\alpha-\beta)\lambda_{(12),(34)},\\\lambda_{(12),3}&=(1-\beta)\lambda_{(12),(34)},\\\lambda_{2,(34)}&=(1-\alpha)\lambda_{(12),(34)},\\2\lambda_{1,(23)}&=(1+3\alpha-\beta)\lambda_{(12),(34)},\\2\lambda_{(23),4}&=(1-\alpha+3\beta)\lambda_{(12),(34)},\\2\lambda_{1,(2(34))}&=(3\alpha+1)\lambda_{(12),(34)},\\2\lambda_{((12)3),4}&=(3\beta+1)\lambda_{(12),(34)},\\4\lambda_{1,((23)4)}&=(3+5\alpha+\beta)\lambda_{(12),(34)},\\4\lambda_{(1(23)),4}&=(3+\alpha+5\beta)\lambda_{(12),(34)}.
\end{align}

For completeness, we also spell out the translation parameters, $\epsilon_i$, given by
\begin{align}
4\epsilon_{1}&=(1+\alpha-\beta)\lambda_{(12),(34)},\\4\epsilon_{2}&=(3+3\alpha+\beta)\lambda_{(12),(34)},\\2\epsilon_{3}&=(2+\alpha+\beta)\lambda_{(12),(34)},\\2\epsilon_{4}&=(1+\beta)\lambda_{(12),(34)}.
\end{align}

The additional conditions in \eqref{eq:convergence-necessary} are related to the fact that the ratios to be Taylor-expanded in $A_5^{(i)}$ are of the form $(1+\Lambda)^S$. Therefore, we need $\Lambda<1$ for convergence.

\end{document}